\documentclass[10pt]{article}
\usepackage{latexsym}
\usepackage{mathptmx}
\usepackage{amssymb,amsmath,graphicx}
\usepackage{iopconf}
\usepackage{amsfonts}
\begin{document}

\newtheorem{lemma}{Lemma}
\newtheorem{proposition}[lemma]{Proposition}
\newtheorem{theorem}[lemma]{Theorem}
\newtheorem{corollary}[lemma]{Corollary}
\newtheorem{assumption}[lemma]{Assumption}
\newtheorem{remark}[lemma]{Remark}

\newcommand{\N}{{\mathbb N}}
\newcommand{\Z}{{\mathbb Z}}
\newcommand{\R}{{\mathbb R}}
\newcommand{\C}{{\mathbb C}}
\newcommand{\A}{\mathbf{A}}
\newcommand{\B}{\mathbf{B}}
\newcommand{\cA}{\mathcal{ A}}
\newcommand{\cB}{\mathcal{ B}}
\newcommand{\cC}{\mathcal{ C}}
\newcommand{\cD}{\mathcal{ D}}
\newcommand{\cE}{\mathcal{ E}}
\newcommand{\cF}{\mathcal{ F}}
\newcommand{\cG}{\mathcal{ G}}
\newcommand{\cH}{\mathcal{ H}}
\newcommand{\cK}{\mathcal{ K}}
\newcommand{\cL}{\mathcal{L}}
\newcommand{\cP}{\mathcal{ P}}
\newcommand{\cQ}{\mathcal{ Q}}
\newcommand{\cS}{\mathcal{ S}}
\newcommand{\sig}{\sigma}
\newcommand{\alp}{\alpha}
\newcommand{\bet}{\beta}
\newcommand{\gam}{\gamma}
\newcommand{\kap}{\kappa}
\newcommand{\lam}{\lambda}
\newcommand{\del}{\delta}
\newcommand{\eps}{\varepsilon}
\newcommand{\ome}{\omega}
\newcommand{\Gam}{\Gamma}
\newcommand{\Lam}{\Lambda}
\newcommand{\Ome}{\Omega}
\newcommand{\lap}{{\Delta}}
\newcommand{\gra}{\nabla}
\newcommand{\degree }{{\rm deg}}
\newcommand{\Dom}{{\rm Dom}}
\newcommand{\Quad}{{\rm Quad}}
\newcommand{\Spec}{{\rm Spec}}
\newcommand{\Ker}{{\rm Ker}}
\newcommand{\Ran}{{\rm Ran}}
\newcommand{\norm}{\Vert}
\renewcommand{\Re}{{\rm Re}\;}
\renewcommand{\Im}{{\rm Im}\;}
\newcommand{\supp}{{\rm supp}}
\newcommand{\Proof}{\underbar{Proof}{\hskip 0.1in}}
\newcommand{\hash}{\#}
\newcommand{\Num}{{\rm Num}}
\newcommand{\conv}{{\rm conv}}
\newcommand{\dist}{{\rm dist}}
\newcommand{\Schrodinger}{Schr\"odinger }
\newcommand{\tnorm}{|\!|\!|}
\newcommand{\dsc}{\mathrm{dsc}}
\newcommand{\ess}{\mathrm{ess}}

\newcommand{\la}{{\langle}}
\newcommand{\ra}{{\rangle}}
\newcommand{\pr}{\prime}
\newcommand{\imp}{\Rightarrow}
\newcommand{\equ}{Leftrightarrow}
\newcommand{\sigl}{\sig_{{\rm loc}}}
\newcommand{\var}{{\rm var}}
\newcommand{\tensor}{\otimes}

\newcommand{\ud}{\mathrm{d}}
\def\[{\bigl[}
\def\]{\bigr]}
\def\({\bigl(}
\def\){\bigr)}
\def\p{\partial}
\def\o{\over}
\def\cm{\cal M}
\def\R{\bf R}
\def\be{\begin{equation}}
\def\ee{\end{equation}}
\def\bea{\begin{eqnarray}}
\def\eea{\end{eqnarray}}
\def\nn{\nonumber}
\def\pd{\partial}
\def\a{\alpha}
\def\b{\beta}
\def\g{\gamma}
\def\d{\delta}
\def\m{\mu}
\def\n{\nu}
\def\t{\tau}
\def\l{\lambda}
\def\L{\Lambda}
\def\s{\sigma}
\def\e{\epsilon}
\def\scri{\mathcal{J}}
\def\cM{\mathcal{M}}
\def\tcM{\tilde{\mathcal{M}}}
\def\RR{\mathbb{R}}
\def\CC{\mathbb{C}}

\begin{flushright}
DFTT-01/2007\\
\end{flushright}
\vspace{1cm}

\title
{Exact discreteness and mass gap of $N=1$ symplectic Yang-Mills  from
M-theory.}
\author {L. Boulton \*\footnote{E-mail: {\tt L.Boulton@ma.hw.ac.uk}}, M. P. Garc\'\i a del Moral\footnote{E-mail: {\tt
 garcia@to.infn.it}}, A. Restuccia\footnote{E-mail:
 {\tt arestu@usb.ve}}}
 \affil{\textsf{$1$}\ Department of Mathematics and
the Maxwell Institute for Mathematical Sciences, Heriot-Watt
University, Edinburgh EH14 2AS, UK} \affil{\textsf{$2$}\
Dipartamento de Fisica Teorica, Universita di Torino and I.N.F.N.,
Sezione di Torino, Via P. Giuria 1, I-10125 , Torino Italy}
\affil{\textsf{$3$}\ Departamento de F\'\i sica, Universidad Sim\'on
Bol\'\i var, Apartado 89000, Caracas 1080-A, Venezuela.}
\beginabstract
In this note we summarize some of the results found recently
in \cite{man1}. 
We show the pure discreteness of the
non-perturbative quantum spectrum of a symplectic Yang-Mills theory
defined on a Riemann surface of positive genus, living in a target
space that, in particular, can be $4$D. This theory corresponds to the membrane
with central charges. The presence of the central charge induces
a confinement in the phase at zero temperature. 
When the energy rises, the center of
the group breaks and the theory enters in a quark-plasma phase after
a topological transition corresponding to the $N=4$ wrapped
supermembrane.
\endabstract

\section{Introduction}
The non-perturbative quantization of String Theory is still an open
problem which receives much of the attention of specialists. 
It can be reformulated in terms of the quantization of
the M-theory in 11 dimensions which, in turns,
reduces to finding the quantization of the basic ingredients: M2-brane or
supermembrane and  M5 brane.
A further important open problem, not necessary connected with
String Theory, is the non-perturbative quantization of Yang-Mills
theories. Attempts along this 
direction include lattice QCD, twistors,
gauge-gravity duality, spin chains, large N matrix models and canonical
quantization.

The aim of this note is to draw to the attention of specialists,
the fact that the membrane with
central charges, which is the quantum equivalent of a symplectic
noncommutative Yang-Mills theory \cite{MOR},\cite{MR}, has a purely discrete spectrum
\textit{at the exact level} of the theory. This is an extension 
of previous results found for the
regularized supermembrane with central charges,
(\cite{M1}-\cite{BR}). 

The correspondence with an $N=1$
Yang-Mills theory defined on a ($2+1$)D Riemann surface of positive
genus $g$ that can live in a target space of $4$D, allows this theory 
to be of interest also outside the scope of String Theory. Admitting an
interpretation in terms of SQCD, it consists of two different
phases at zero temperature: a confined phase comprising glueballs in
the bosonic sector and a quark-gluon plasma phase which has a microscopic
origin in the M-theory. The quantum consistency of supermembranes with
fixed central charges provides then an indirect proof of consistency
of all these noncommutative gauge theories.

\section{The supermembrane with non-trivial central
charge} Supermembranes are extended objects defined in terms of a
base manifold, a Riemann surface $\Sigma$, which live in a
Minkowski target space. The canonically reduced Hamiltonian in the
light cone gauge \cite{DWHN} has the expression
\begin{equation}
   \int_\Sigma  \sqrt{W} \left(\frac{1}{2}
\left(\frac{P_M}{\sqrt{W}}\right)^2 +\frac{1}{4} \{X^M,X^N\}^2+
{\small\mathrm{\ Fermionic\ terms\ }}\right)
\end{equation}
where 
\be \{X^{M}, X^{N}\}^{2}= \frac{\epsilon
^{ab}}{\sqrt{W(\sigma)}}\partial_{a}X^{M}\partial_{b}X^{N}. 
\ee 
restricted by the first class constraint,
\begin{equation} \label{e1}
   \oint_\mathcal{C} \frac{P_M}{\sqrt{W}} \quad X^M = 0.
\end{equation}
which generates area preserving diffeomorphisms of $\Sigma$ for $\mathcal{C}$ any given closed path. Here and below $M,N=1,\ldots,9$.
The continuity of the spectrum of the above  Hamiltonian at the $SU(N)$ regularized level was demonstrated in
\cite{DWLN}. This
property relies on two basic facts: supersymmetry and the presence of singular
configurations with zero energy at a classical level. 

Under compactifications, for example regard the target space as being
$M_{10}\times
S^1$, it is believed, \cite{DWPP},
that the spectrum of the theory remains continuous.
Therefore, the compactification
procedure by itself, does not seem enough to change the spectral 
properties of the model. 

We now impose some topological restrictions on the configuration
space. These completely characterize the $D=11$ supermembrane with
non-trivial central charge generated by the wrapping on the compact
sector of the target space. We will assume 
that the target space is $M_{9}\times S^{1}\times S^{1}$and its base manifold $\Sigma$ of positive genus $g$,
 for simplicity, $g=1$. We stress that
these properties remain valid for any other target space of
dimension less than $9$ in particular $D=4$. All maps from the base space
$\Sigma$, must satisfy
\begin{equation} \label{e2}
\begin{aligned}
  &\oint_{C_i} \ud X^r=2\pi S^r_iR^r, \quad &r=1,2 ; \qquad
  &\oint_{C_i} \ud X^m=0 \quad &m=3,\ldots,9
\end{aligned}
\end{equation}
for $i=1,2$ and the topological condition,
\begin{equation} \label{e3}
  Z= \int_\Sigma \ud X^r\wedge\ud X^s =\epsilon^{rs}(2\pi^2R_1R_2)n,
\end{equation}
 where $n=\det S^{r}_i$ is fixed, each entry $S^r_i$
is integer, and $R_1$ and $R_2$ denote the radii of the target
component $S^1\times S^1$. Note that (\ref{e2}) describe maps from
$\Sigma$ to $S^1\times S^1$ with $\ud X^m$ a non-trivial closed
one-form. The only restriction upon these maps is the assumption
that $n$ is fixed. The term on the left side of (\ref{e3}) describes
the central charge of the supersymmetric algebra. Among all the maps 
from the torus $\Sigma$ to the target space satisfying (\ref{e2}),(\ref{e3}) there is a minimizer of the Hamiltonian. It corresponds to
a minimal immersion from $\Sigma$ to the target space which implies
that, for the case of flat target spaces, the worldvolume of the
supermembrane is a calibrated submanifold.

 Minimal immersions can
also describe non-BPS minimal solutions, \cite{JR}.
The theory results to be invariant under $SL(2,Z)$. The
degrees of freedom are expressed in terms of $A_{r}$ and the
discrete set of integers described by the harmonic one-forms. We can
always fix these gauge transformations by
\begin{equation}
   S^r_i=l^r \delta^r_i, \quad l^1 l^2=n.\quad\to\quad
   \ud X^r=2\pi R^r l^r \ud \hat{X}^r + \delta^r_s \ud A_s.
\end{equation}
After the gauge fixing there is a residual transformation \cite{man1}
$\mathbb{Z}(2)$ under which,
\begin{equation}
   A_{1}\to A_{2}\quad A_{2}\to -A_{1}.
\end{equation}
This allows us to rewrite the Hamiltonian in terms of $X^{m}, m=1,..,7$
and $A_{r}, r=1,2$. The resulting expression is:
\begin{equation}\label{e5}
 \begin{aligned}
H&=\int_{\Sigma}\frac{1}{2}\sqrt{W}[P^{2}_{m}+ \Pi^{2}_{r}+
\frac{1}{2}W\{X^{m},X^{n}\}^{2}+W(\mathcal{D}_{r}X^{m})^{2}+\frac{1}{2}W(\mathcal{F}_{rs})^{2}]
\\&+\int_{\Sigma}[\frac{1}{8}\sqrt{W}n^{2}-\Lambda
(\mathcal{D}_{r}\Pi_{r}+\{X^{m},P_{m}\})]\\& + \int_{\Sigma}
\sqrt{W} [- \overline{\Psi}\Gamma_{-} \Gamma_{r}
\mathcal{D}_{r}\Psi +\overline{\Psi}\Gamma_{-}
\Gamma_{m}\{X^{m},\Psi\}+
 \Lambda \{ \overline{\Psi}\Gamma_{-},\Psi\}]
 \end{aligned}
 \end{equation}
where (\cite{MOR}, \cite{M1})
$\mathcal{D}_r\bullet=2D_{r}\bullet +[A_{r},\bullet]$,
$\mathcal{F}_{rs}=\mathcal{D}_rA_s-\mathcal{D}_s A_r+ [A_r,A_s]$,
and $P_{m}$ and $\Pi_{r}$ are the conjugate momenta
to $X^{m}$ and $A_{r}$ respectively. $\mathcal{D}_{r}$ and
$\mathcal{F}_{rs}$ are the covariant derivative and curvature of a
symplectic noncommutative theory \cite{MOR,M2}, constructed from the
symplectic structure $\frac{\epsilon^{ab}}{\sqrt{W}}$ introduced by
the central charge. The last term represents its supersymmetric
extension in terms of Majorana spinors. The relevant degrees of
freedom in order to quantize the theory are the $X^{m}$ and the gauge invariant part of $ A_{r}$ which
are single valued over the base manifold. A $SU(N)$ regularization was
obtained in \cite{M1} and the spectral properties of the spectrum
were rigourously demonstrated at classical level, and at quantum level in
several papers, \cite{M2},\cite{M3}, \cite{BR} .

\section{On the spectrum of the exact theory}
According to the results reported in \cite{M2}, the bosonic
regularized Hamiltonian of the $D=11$ supermembrane with central
charge, $H_{N}^{B}$, relates to its semi-classical approximation,
$H_{\mathrm{sc},N}^{B}$, by means of the fo\-llo\-wing operator
inequality:
\begin{align} \label{fi3}
H_{N}^{B}\geq C_{N}H_{\mathrm{sc},N}^{B}.
\end{align}
Here $N$ denotes the size of the truncation in the Fourier basis of
$\Sigma$ and $C_N$ is a positive constant. A seemingly crucial step
in the proof of \eqref{fi3} found in \cite{M2}, relies heavily on
the compactness of the unit ball of the configuration space which
happens to be finite dimensional. The exact bosonic Hamiltonian
however contains a configuration space which is
\textit{infinite}-dimensional, so that the unit ball is not compact.
In \cite{man1} we show that the same operator relation holds
true for the exact bosonic Hamiltonians. We overcome the difficulty
of the analysis by carrying out a detailed analysis of each term
involved in the expansion of the potential term of the exact 
bosonic Hamiltonian, $H^B$.

Following the standard notation $L^p\equiv
L^p(\sigma)$ denotes the Banach space of all fields $u$, such that $
\|u\|_{p}=\langle u ^{p} \rangle^{1/p}<\infty.$ Let
\begin{align} \label{fi7}
\|u\|_{4,2}=(\|D_{r}u\|^4+\|D_rD_s u\|^4)^{1/4}.
\end{align}
The fields $X^m$, $A_r$ lie on
the \emph{configuration space} $\mathcal{H}^{4,2}$ of functions
$u\in \mathcal{H}^1$ such that $\|u\|_{4,2}<\infty$. Note that the
left hand side of \eqref{fi7} is a well defined norm in
$\mathcal{H}^{4,2}$, the later is a linear space, but we do not make
any assumption about completeness.
The potential, $V$,  of the bosonic sector of the supermembrane with
central charges is well defined in $\mathcal{H}^{4,2}$ as
\begin{align} \label{fi1}
V=\langle
\mathcal{D}_{r}X^{m}\mathcal{D}_{r}X^{m}+\frac{1}{4}\mathcal{F}_{rs}
\mathcal{F}_{rs} \rangle.
\end{align}
 $V$ is not well defined in $\mathcal{H}^1$
but in $\mathcal{H}^{4,2}$. 

By imposing the gauge fixing
conditions, $D_{1}A_{1}=0$ and $D_{1}A_{2}=0$, the potential can be
re-written as,
\begin{equation}
V=\rho^{2}+2\B+\A^2
\end{equation}
where $\rho^{2}$ is the semiclassical potential term of
$H^B_{\mathrm{sc}}$,
Let $\rho^{2}$ be the potential term of
$H^B_{\mathrm{sc}}$, so that
\begin{align*}
\rho^{2}=\langle D_{r}X^{m}D_{r}X^{m}+
(D_{1}A_{2})^{2}+(D_{2}A_{1})^{2} \rangle.
\end{align*} and

\begin{gather}
\B=\langle D_rX^m\{A_r,X^m\}+D_1A_2\{A_1,A_2\} \rangle \\
\A=\langle \{A_1,X^m\}^2+\{A_2,X^m\}^2+\{A_1,A_2\}^2+\{X^m,X^n\}^2
\rangle.
\end{gather}
This allows us to show the following crucial identity, \cite{man1}:
there exists a constant $0<C\leq 1$, such that
\begin{equation} \label{fi4}
  V\geq C \rho ^2, \qquad \qquad \forall X^m,\, A_r \in \mathcal{H}^2.
\end{equation}
The latter is a consequence of 
the particular expression of the potential, properties as the
compactness of the base manifold (not of the configuration space) and
the Cauchy-Schwarz inequality. 

In order to define rigorously the Laplacian  in the
non-compact infinite dimensional configuration space we have
introduced, the Hamiltonian is expressed as
\begin{align*}
  H^{B}=[V_{quartic}+V_{cubic}+(1-C)V_{quadratic}]+[-\Delta+C V_{quadratic}]
\end{align*}
where the first bracket acts multiplicatively on the Hilbert space
of states, while the operator on the second bracket may be expressed
in terms of creation and annihilation operators in the usual way.
Alongside with (\ref{fi4}), this expression ensures the operator
identity 
\begin{align}
H^{B}\geq CH_{\mathrm{sc}}^{B}.
\end{align}
analogous to (\ref{fi3}).

\section{Confinement of the theory}
It was an original idea of G. 't Hooft, \cite{thooft} that
permanent quark confinement occurs in a gauge theory if its vacuum
condenses into a state which resembles a superconductor. His
proposal was to consider the confinement of quarks as dual of the
Meissner effect, where the role of magnetism and electricity are
interchanged. In his approach he considered a nonabelian gauge
theory that were seen as an abelian theory enriched with Dirac
magnetic monopoles, see also \cite{witten}. The symplectic Yang-Mills naturally creates this
effect. The mass contribution of the central charge, or,
analogously, its correlated residual $Z(2)$ symmetry  of the
Hamiltonian, can be described in terms of the quadratic derivatives
of the configuration fields $X^{m}$ and $A_{r}$. The derivatives of
these fields correspond to mappings of the target space into
$\Sigma$. These are induced
 by the minimal immersion which realises by 
$\widehat{X}_{r}$, $r=1,2$, and the harmonic fields over $\Sigma$,
\begin{align*}
D_{r}Y_{A}=\{\widehat{X}_{r},Y_{A}\}=\lambda_{rA}^{B}Y_{B}=\lambda_{rA}Y_{A}
\end{align*}
where
\begin{align*}
\lambda_{rA}^{B}=\int
d^{2}\sigma\sqrt{\omega}\{\widehat{X}_{r},Y_{A}\}Y^{B}.
\end{align*}
They correspond to a particular subset of the structure constants
that mixes the harmonic and the exact forms, $g_{rA}^{C}$. 

For the
case of a torus, an explicit relation was found in \cite{M1}. The
quadratic terms on the derivatives of the configuration variables,
define a strictly positive function whose contribution to the
overall Hamiltonian gives rise to a basin shaped potential. The
latter eliminates the string-like spikes and provides a discrete
spectrum, even for the supersymmetric model. The centre created by a
discrete symmetry is a mechanism for providing mass to the
monopoles \cite{pepe}.  The supermembrane theory when compactified in $4$D can
be interpreted as a theory modelling  susy QCD in the spirit of \cite{chodos}. It exhibits
confinement in the phase at zero temperature since the theory
becomes naturally the supermembrane with central charges, which has
minimal energy. 

The mass terms are determined by the elements of
the centre $m(z)$ associated to the correlation length of the
particles \cite{pepe2}. 
By rising the energy, the
theory enter in the phase of asymptotic freedom described by the
supermembrane without central charges. The phase transition is
described by the breaking of the center of the group that becomes
trivial. The phase transition of topological nature \cite{preskill}-\cite{balachandran}. 
The particles behave as if they where in a quark-gluon
plasma. These quarks-gluons do not feel the
topological effects, since the correlation length becomes infinite
and the effective volume is zero. Along the commutative directions
the quarks experiment no force.

 \section{Conclusions}
In this note we have sumarized the results of the recent manuscript
\cite{man1}. The
$D=11$ supermembrane with central charges is quantum equivalent to
the $N=1$ 2+1 Symplectic non-commutative Super Yang-Mills Theory
defined in target spaces of dimension $D\le 9$. The spectrum of the
bosonic sector of the $D=11$ theory has been demonstrated at exact level
of the theory to be purely discrete, hence containing a mass gap.
The theory exhibit confinement in the supermembrane with central
charge phase. It enters in the asymptotic free phase
through the spontaneous breaking of the center. This phase
corresponds to the $N=4$ wrapped supermembrane.

\section*{Acknowledgements}
We thank M. Banados, J. Zanelli, for helpful
conversations. M.P.G.M thanks the organizers of the Workshop ``'ForcesUniverse 2006'' 
for the opportunity to present this work, and  to CECS, (Valdivia, Chile) for kind
hospitality and support. M.P.G.M. is partially supported by the
European Comunity's Human Potential Programme under contract
MRTN-CT-2004-005104 and by the Italian MUR under contracts
PRIN-2005023102 and PRIN-2005024045.

\end{document}